\documentclass[12pt]{iopart}
\usepackage{graphicx}
\usepackage{epsfig}

\begin{document}

\title[Drilled HTS: how to arrange the holes to maximize the trapped magnetic flux ?]{Bulk high-Tc superconductors with drilled holes: how to arrange the holes to maximize the trapped magnetic flux ?}
\author{Gregory P~Lousberg$^{1,3}$, M~Ausloos$^{2}$, Ph~Vanderbemden$^{1}$, and B~Vanderheyden$^{1}$}
\address{$^1$ SUPRATECS Research Group, Dept. of Electrical Engineering and Computer Science (B28), University of Li\`ege, Belgium}
\address{$^2$ SUPRATECS, Dept. of Physics (B5), University of Li\`ege, Belgium}
\address{$^3$ FRS-FNRS fellowship}
\ead{gregory.lousberg@ulg.ac.be}

\begin{abstract}
Drilling holes in a bulk high-Tc superconductor enhances the oxygen annealing and the heat exchange with the cooling liquid. However, drilling holes also reduces the amount of magnetic flux that can be trapped in the sample.  In this paper, we use the Bean model to study the magnetization and the current line distribution in drilled samples, as a function of the hole positions. A single hole perturbs the critical current flow over an extended region that is bounded by a discontinuity line, where the direction of the current density changes abruptly. We demonstrate that the trapped magnetic flux is maximized if the center of each hole is positioned on one of the discontinuity lines produced by the neighbouring holes. For a cylindrical sample,  we construct a polar triangular hole pattern that exploits this principle; in such a lattice, the trapped field is $\sim20\%$ higher than in a squared lattice, for which the holes do not lie on discontinuity lines. This result indicates that one can simultaneously enhance the oxygen annealing, the heat transfer, and maximize the trapped field.
\end{abstract}

\submitto{\SUST}


\maketitle

\section{Introduction}

High-temperature bulk superconductors are very promising materials for permanent magnet applications~\cite{1,2,3,4,Supratecs}. They can be used in magnetic bearings (in the Maglev train~\cite{5} or in frictionless linear translation systems~\cite{6}) and in rotating machines (synchronous motors~\cite{motor,7} or flywheels for energy storage~\cite{flywheel,8}). At the liquid nitrogen temperature, such magnets are able to trap up to 3~T~\cite{Murakami}. When cooled down to 29~K, the maximum trapped field can reach 17~T~\cite{9}. 

Recently, it has been proposed to drill arrays of columnar holes inside high-Tc superconducting magnets in order to improve their chemical and thermal properties~\cite{FirstDrilled,10}. First, the holes reduce the oxygen diffusion wall and enhance the oxygen annealing process~\cite{11}. Second, the larger exchange surface increases the heat transfer with the environment and is thus beneficial for the cooling of the superconductor~\cite{12}. A rapid cooling is required for instance when a superconductor is magnetized with a pulsed field \cite{13}, because the dissipative motion of vortices tends to raise rapidly the temperature of the material and thus to reduce both the critical current density and the trapped magnetic flux. A third (although counterintuitive) advantage to drilling holes in a superconductor is to improve their mechanical properties. Samples can be strengthened by impregnating the holes with a reinforcement resin that prevents cracks from developing~\cite{14}, for instance as a result of strains induced by the Lorentz force~\cite{9}. 

Drilling holes in a superconductor is however detrimental to its magnetic properties. It was found in~\cite{15,16} that removing superconducting matter decreases both the full penetration field and the trapped flux. Holes also lead to macroscopic changes in the current distribution. In the Bean model, the current stream lines near a hole abruptly change their direction along discontinuity lines~\cite{18} and circle the hole in a region that extends far beyond the hole itself. This effect is enhanced in thin films, as the magnetic flux density displays sharp peaks at the discontinuity lines. Such macroscopic changes of the magnetic flux were observed with magneto-optical imaging of thin films with macroscopic defects~\cite{19,20}. For bulk samples, studies based on the Bean model already pointed to the magnetization drop that results from drilling holes~\cite{15}. It was also shown that for a given lattice, the magnetization drop increases with the diameter of the holes~\cite{15,16}. It has been measured in~\cite{16} that increasing the hole diameter by a factor of 2 results in a magnetization drop of $\sim~80\%$. In the particular limit of YBCO thin films of rectangular shape with microscopic holes, the Bean critical state has also been simulated in~\cite{Crisan}. However, to our knowledge, none of these previous works has studied the influence of the hole \emph{pattern} on the magnetization drop. 

In this paper, we investigate the effect of the arrangement of holes on the magnetization drop of drilled samples, by studying the current distribution and the interaction among the influence regions of the holes. For that purpose, we develop an algorithm based on the Bean model and on an observation made by Campbell and Evetts~\cite{18} to calculate the magnetic field in the critical state for an infinitely long drilled sample with an arbitrary cross section. 

This paper is organized as follows. The algorithm is discussed in section~\ref{s:model} and is used in section~\ref{s:one} to calculate the magnetization of a sample with a semi-infinite cross section and a single hole. In section~\ref{s:several}, we study the magnetization of samples with either two or three holes, as a function of their relative positions. Section~\ref{s:lattice} is devoted to the magnetization drop in samples with either a semi-infinite or a circular cross section and holes disposed on a lattice pattern. Section~\ref{s:conclusions} concludes this work. 

\section{Model for the magnetic field distribution in drilled samples }
\label{s:model}

\begin{figure}[t]
\center
\includegraphics[width=7cm]{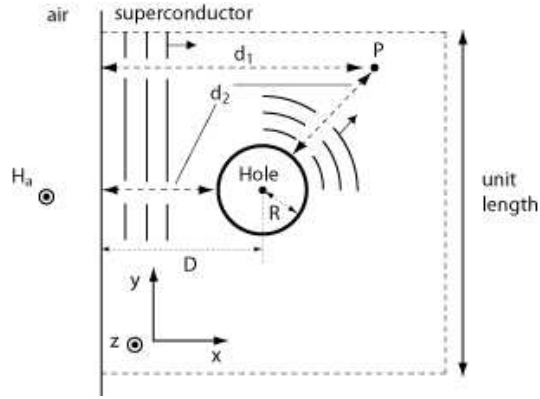}\caption{Sketch of an infinitely long sample with a semi-infinite cross section drilled by a single hole of radius $R$ located at a distance $D$ from the border. The flux front can reach the point $P$ by following two paths: directly from the border, with a path length $d_1$, or via the hole, which acts as a radial source, yielding a path length $d_2$.}\label{geom}
\end{figure}

In this paper, we neglect demagnetization effects and focus on superconducting samples that are infinitely long and have either a semi-infinite or a circular cross section. Figure \ref{geom} shows a sample with a semi-infinite cross section containing a circular hole of radius $R$ located at a distance $D$ from the border. The cross section lies in the $x-y$ plane. We further assume that the applied magnetic field $H_a$ is oriented along the $z$-axis and is uniform. Its amplitude is such that $H_{c1} \ll H_a \ll H_{c2}$. We assume strong pinning and neglect surface barrier effects. Under these assumptions, the distribution of the magnetic field in the sample cross section is described by the Bean model \cite{17}, which gives
\begin{equation}
\label{Bean} \frac{dH_z}{d\ell}(P)=\left\{\begin{array}{l}0\\ \pm J_c\end{array}\right.
\end{equation}
where $J_c$ is constant, while $\ell$ represents the  distance traveled by the flux front to reach a given point, $P$. 

Let us illustrate the procedure to determine the field distribution in the example shown in Figure~\ref{geom}. For a point $P$ located at a distance $d_1 > D$, the flux front can travel along two paths: it can reach $P$ directly from the border, with a path length $d_1$, or via the hole which acts as a \textit{radial source} of magnetic field \cite{18}, with a total path length $d_2$. Following Campbell and Evetts~\cite{18}, we assume that the flux front travels along the shortest path. Hence, the points in a flux front are located at a fixed length $\ell = H_a/J_c$ from the border, where $\ell$ is evaluated as $d_1$ or $d_2$, whichever is smaller. The magnetic field, $H=H_z$, at a given point, $P$, can then be calculated by determining the length, $\ell$, of the shortest path that reaches $P$ and by evaluating 
  \begin{eqnarray}
    H = H_a - J_c \,\ell.
  \end{eqnarray}
Once the distribution of $H$ is known, the current stream lines can also be easily obtained, as they coincide with the lines of constant magnetic field. Finally, the magnetization of the sample is given by
\begin{eqnarray}
  M=\frac{1}{\mu_0 S}\int B \,dS-H_a = \frac{1}{S}\int H \,dS-H_a
\end{eqnarray}
where $S$ is the sample cross section (for a semi-infinite cross section, $S$ is arbitrarily chosen to be a square section of unit length) and we assume $B = \mu_0 H$. 

\begin{figure}[t]
\center
\includegraphics[width=7cm]{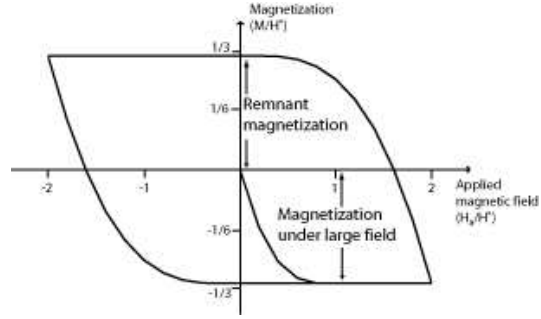}\caption{Simulated magnetization curve of an infinitely long sample with a circular cross section (radius $a$) containing one hole of radius $R=0.05~a$ located at a distance $D=0.2~a$ from the border.  The applied field, $H_a$, first increases from $0$ to $2H^*$, where $H^*$ is the penetration field,  then decreases from $2H^*$ to $-2H^*$, and finally increases again to $H^*$. The magnetization obtained for $H_a > H^*$ is equal in magnitude to the remnant magnetization.}\label{Magn}
\end{figure}

We use these principles to study samples with an arbitrary number of holes and construct an algorithm that calculates the magnetic field distribution as a function of the hole radii and positions. In the following sections, we address two questions: (i) what is the magnetization of a given sample that is subjected in the zero field cooled state to an applied field, $H_a$, and (ii), what is the remnant magnetization that is obtained when the same sample is first magnetized above twice its penetration field before the applied field returns to zero. For samples with a finite cross section, both magnetizations actually have the same magnitude (they have opposite signs, however), provided that the applied field in situation (i) is larger than the penetration field. Figure~\ref{Magn} illustrates this equivalence for the case of a sample with a circular cross section containing one hole, a case which will be treated in section~\ref{s:lattice}. Such equivalence cannot be found for samples with a semi-infinite cross section, as they are never fully penetrated. We will nevertheless consider these systems when subjected to an increasing field (case (i)), because these situations allow us to understand the interaction between different holes.

\section{Samples with one hole}
\label{s:one}
\subsection{Current lines}

\begin{figure}[b]
\center
\includegraphics[width=7cm]{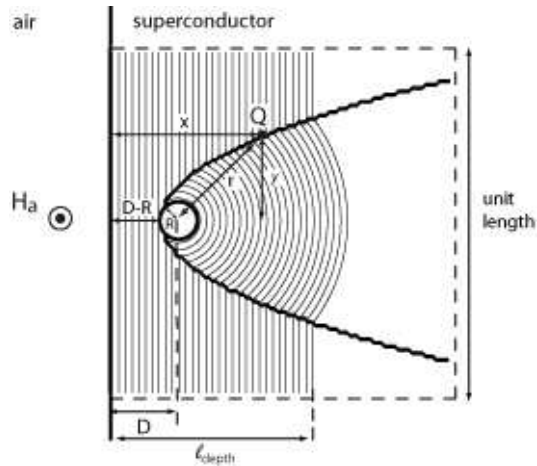}\caption{Simulated current lines (or constant magnetic field lines) in a sample with a semi-infinite cross section and a single hole. The hole has a radius $R=0.05$ and it is located at distance $D=0.2$ from the border. Here, the unit length corresponds to the length of one of the segments of the dashed contour. The applied magnetic field is such that the penetration depth is given as $\ell_\mathrm{depth}=0.6$. For a point $Q$ on the discontinuity line, the path lengths of both penetration routes, $x$ and $D-2R+r$, are equal.}\label{currentline}
\end{figure}

We first consider a sample with a semi-infinite cross section drilled by a single hole of radius $R$ located at a distance $D$ from the border, and apply a magnetic field $H_a$ in the zero field cooled state. Following the main principles of our algorithm, we know that the magnetic field can reach a given point by two distinct penetration routes. We can thus identify two regions: one for which the direct penetration from the border has the shortest path, and one for which the radial penetration via the hole has the shortest path. Hence, the boundary between these regions is characterized by the equality of path lengths,
\begin{equation}
\label{parabola}
x = D + r - 2 R
\end{equation}
where $x$ is a cartesian coordinate along an axis that is
perpendicular to the external boundary and $r$ is the distance from the hole center to the point where we determine the path lengths (see Figure~\ref{currentline}). We thus find that the boundary defined in~(\ref{parabola}) is the locus of points for which the difference between the distance to the external boundary and that to the hole center is equal to a constant, $D-2 R$. This locus is a parabola whose vertex is located at $(x,y)=(O,D-R)$, whose directrix runs along $y=D-2R$, and whose focus lies at $(0,D)$. In cartesian cooordinates, the parabola equation reads
\begin{equation}
\label{parabola2}
x=\frac{y^2}{4R}+D-R.
\end{equation}
It is plotted as a thick line in Figure~\ref{currentline} for the case of a hole of radius $R=0.05$ located at a distance $D=0.2$ from the border. Here, the unit length corresponds to the length of one of the sides of the square delimited by the dashed contour; all the distances are normalized to this length.

Equation~(\ref{parabola2}) also characterizes the current discontinuity line. As explained in the previous section, the current stream lines can be constructed from the contour lines of constant magnetic field. These lines follow straight segments outside the parabola, where the distance to the border is the shortest, and arcs of circle inside the parabola, where the penetration path through the hole is the shortest. The current lines abruptly change their direction on the parabola, which is thus a discontinuity line. Figure~\ref{currentline} shows the current lines obtained when the field is applied in the zero field cooled state and is raised to a finite $H_a$. In the particular case shown, the applied field corresponds to a penetration length $\ell_{\mathrm{depth}} = 0.6$. 

\subsection{Influence of the hole radius on the magnetization drop}

In the case of a sample containing one single hole, the magnetization can be calculated in two ways: either numerically, by using the algorithm described in section~\ref{s:model}, or analytically, by calculating the magnetic flux inside and outside the parabola of Equation~(\ref{parabola2}). The relative magnetization drop incurred by the drilled sample is then given by
\begin{eqnarray}
  \frac{\Delta M}{|M_0|} = \frac{|M_0-M_1|}{|M_0|},
\end{eqnarray}
where $M_0$ is the magnetization of a sample without a hole and $M_1$ is that for a sample with a single hole. The calculations are carried over a square of unit side length. In the particular case considered, the hole center is located at $x=0.2$ and $y=0$, and we let the magnetic field penetrate up to a length $\ell_\mathrm{depth}=0.6$. In units of $J_c$, the applied field is thus given as $H_a = J_c \ell_\mathrm{depth} = 0.6~J_c$. These choices guarantee that the flux does not extend further than $x=1$ in the hole influence region.

\begin{figure}[t]
\center
\includegraphics[width=7cm]{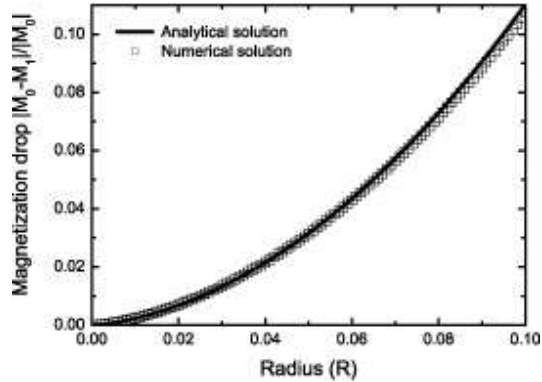}\caption{Magnetization drop in a sample with one hole and a semi-infinite cross section as a function of the hole radius. The hole is located at a distance D=0.2 from the border. The sample is limited to a unit surface for the calculation of the magnetization. Here, $M_0$ is the magnetization of a sample with the same geometry and no hole. The applied magnetic field is $H_a=0.6~J_c$. The solid line corresponds to the analytical solution, and the squared symbols to that obtained with the numerical algorithm.}\label{Analyt}
\end{figure}

Figure~\ref{Analyt} shows the relative magnetization drop, $\Delta M/|M_0|$, as a function of the hole radius $R$ (the analytical result is shown with solid lines, the numerical one is plotted with  square symbols). The detailed analytical calculations are given in~\ref{appendix}. Analytical and numerical calculations are in good agreement. We observe that $\Delta M/|M_0|$ increases with the radius of the hole, as expected intuitively and illustrated in Hall probe mapping experiments~\cite{15}. As shown in~\ref{appendix}, a series expansion of the analytical result for the magnetization drop around $R = 0$ yields
\begin{eqnarray}
\label{M01}
\frac{\Delta M}{|M_0|} &=& \frac{J_c}{|M_0|S} \left(\frac{32}{9}\sqrt{R^3(\ell_\mathrm{depth}-D)^3}+O(R^{5/2})\right), \\
&=& 0.97 \, \left(\frac{R}{\ell_{\mathrm{depth}}}\right)^{3/2} +O\left(\frac{R}{\ell_{\mathrm{depth}}}\right)^{5/2}.
\end{eqnarray}
This is not a trivial result! One could have naively expected that the magnetization drop roughly scales either as the area of the hole, $\Delta M\propto R^2$, or as the area of the region delimited by the parabola, $\Delta M \propto \sqrt{R}$. From Equation~(\ref{M01}), we conclude that an intermediate situation occurs.

\section{Samples with several holes}
\label{s:several}
\subsection{Samples with two holes}

Consider now a sample with a semi-infinite cross section and two holes. Both holes have the same radius $R=0.05$ and are separated by a constant distance $d=0.2$. The first hole is placed at a distance $D=0.2$ from the border. Again, the semi-infinite surface is limited to a square whose sides have a unit length. Let us vary the angular separation $\theta$ between the holes and study the interactions between the regions of influence of the holes; their interaction should depend on whether the center of the second hole lies inside or outside the parabolic influence region of the first hole. We work in polar coordinate and take the origin at the center of the first hole. The coordinates of the second hole are given by $(r,\theta)$ with $r=d$. The discontinuity line produced by the first hole has the equation
\begin{equation}
\label{parabola3}
r=\frac{2R}{1-\sin\theta}.
\end{equation}
Thus, the center of the second hole lies on the parabola when $r = 0.2$, and hence when $\theta = 30^\circ$.

\begin{figure}[t]
\center
\includegraphics[width=10cm]{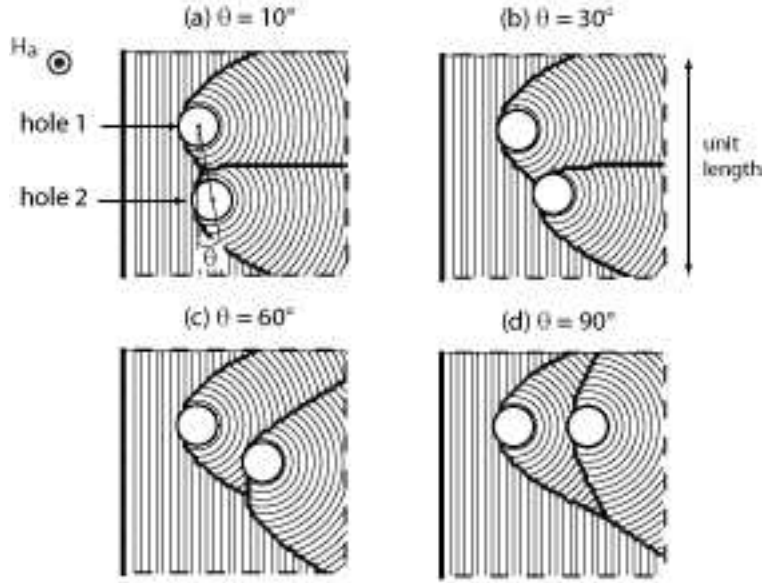}\caption{Simulation of the current lines in a sample with a semi-infinite cross section and with two holes. The holes have a radius $R=0.05$. The unit length corresponds to the side of the square delimited by the dashed lines. The first hole is located at $D=0.2$ from the border. The separation distance between the holes is constant, $d=0.2$. The second hole is located at $\theta=10^\circ$ (a), $\theta=30^\circ$ (b), $\theta=60^\circ$ (c) and $\theta=90^\circ$ (d). The thick lines represent the discontinuity lines. }\label{hyperbole}
\end{figure}

Figure \ref{hyperbole} shows the current lines for four different angular positions $\theta$. For $\theta=10^\circ$, the center of the second hole is located outside the influence region of the first hole. A new discontinuity parabola appears around the second hole. The two parabolic curves merge between the holes and form a common discontinuity line. This last line corresponds to the locus of points for which the difference between the distances to each hole center is equal to a constant; the discontinuity line is therefore a branch of a hyperbola. When $\theta$ increases further, the second hole is pushed away from the border and, for $\theta =30^\circ$, enters the influence region of the first hole. Again, each hole produces a parabolic discontinuity line and the two lines merge into a branch of hyperbola. As the second hole goes deeper in the region of influence of the first one, the hyperbola opens up. The surface of the combined region of influence of the holes increases with $\theta$ and reaches a maximum for $\theta=90^\circ$.

\begin{figure}[t]
\center
\includegraphics[width=7cm]{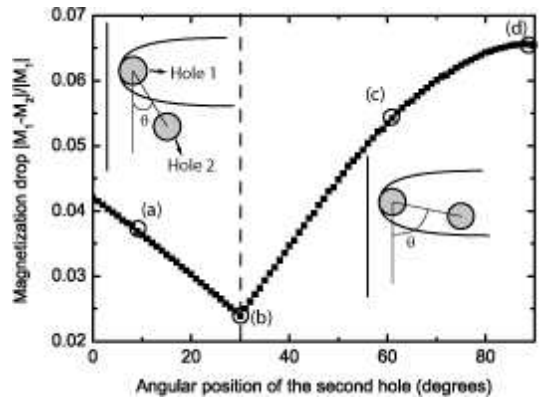}\caption{Magnetization drop in a sample with two holes and a semi-infinite cross section, as a function of the relative angular position of the holes. The first hole has a radius $R=0.05$ and is located at a distance $D=0.2$ from the border. Here, the reference magnetization is calculated on a sample with the same geometry, but containing only the first hole. The second hole has a radius $R=0.05$  and is located at a constant distance from the first hole ($d=0.2$). The applied magnetic field is given as $H_a=0.6~J_c$. The angular positions (a), (b), (c) and (d) correspond to the situations in Figure~\ref{hyperbole}. We observe that $\Delta M/|M_1|$ has a minimum when the center of the second hole lies on the discontinuity line of the first hole.}\label{Twoholes}
\end{figure}

We evaluated the magnetization drop induced by the second hole as
\begin{eqnarray}
\frac{\Delta M}{|M_1|}=\frac{|M_1-M_2|}{|M_1|},
\end{eqnarray}
where $M_1$ is the magnetization for the sample with hole $1$ only, and $M_2$ is that for the sample with holes $1$ and $2$. We can in principle evaluate this expression either by following the numerical method exposed in section~\ref{s:model}, or analytically. However, analytical calculations rapidly become tedious when several holes are involved; we will thus restrict ourselves to numerical results from now on. The magnetization drop is plotted in Figure~\ref{Twoholes} as a function of the relative angular position of the holes.  For small $\theta$, the magnetization drop decays as the angle is increased. This result follows from the fact that the second hole is pushed away from the border as $\theta$ increases; the hole is thus threaded by a lower magnetic flux and its effect is reduced. By contrast, for large angles, $\Delta M/|M_1|$ increases with $\theta$ because the influence region of the second hole becomes larger.  Hence, the optimal position of the center of the second hole is right on the discontinuity line of the first hole:  the magnetization drop is minimum for $\theta=\theta_{\mathrm{opt}}=30^\circ$. Although the results are not shown, we have also studied the situation with a fixed relative angular position and a variable separation distance and found similar conclusions.

\begin{figure}[t]
\center
\includegraphics[width=7cm]{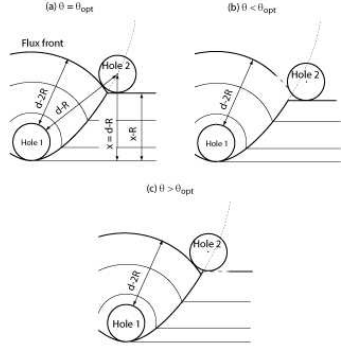}\caption{Sketch of the flux front tangent to hole 2 for angular positions $\theta=\theta_{\mathrm{opt}}$ (a), $\theta<\theta_{\mathrm{opt}}$ (b), and $\theta>\theta_{\mathrm{opt}}$ (c). The dashed parts are the remnant flux front in the influence region of the second hole. When the center of the second hole is located on the discontinuity parabola, the flux front is tangent to the second hole simultaneously in the regions inside and outside the parabola.}\label{optimum}
\end{figure}

Some insight on these results can be gained by examining how the flux penetrates the system. The flux front near the second hole is sketched  in Figure~\ref{optimum} for  $\theta=\theta_{\mathrm{opt}}=30^\circ$ (a), $\theta<\theta_{\mathrm{opt}}$ (b), and $\theta>\theta_{\mathrm{opt}}$ (c). We can observe that the flux front reaches the second hole tangentially in all cases. However, for $\theta=\theta_{\mathrm{opt}}$, the flux front is tangent to the second hole simultaneously in the region inside the discontinuity parabola produced by the first hole (circular front) and in the region outside the parabola (straight front). The simultaneous penetration from the two regions appear to be necessary for reducing the effect of the second hole on the magnetization of the sample.

\subsection{Samples with three holes}

We now turn to adding a third hole to the optimized two-hole pattern of Figure~\ref{optimum}-(a), where the first hole is again located at $D=0.2$ away from the external border and the center of the second hole lies on the discontinuity parabola of the first one, at a distance $d=0.2$. The three holes have a radius $R=0.05$. The third hole is located at a constant distance $d = 0.2$  from the second hole. Adding a third hole reduces again the magnetization of the sample.  The magnetization drop is now given by
\begin{eqnarray}
\frac{\Delta M}{|M_2|}=\frac{|M_2-M_3|}{|M_2|},  
\end{eqnarray}
where $M_2$ is the magnetization of a sample containing only the first two holes, and $M_3$ is the magnetization for a sample containing three holes. 

\begin{figure}[t]
\center
\includegraphics[width=7cm]{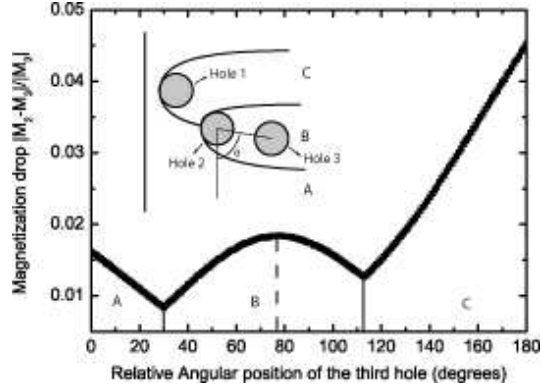}\caption{Magnetization drop of a sample with three holes and a semi-infinite cross section, as a function of the relative angular position between the second and the third hole. The first hole has a radius $R=0.05$ and is located at a distance $D=0.2$ from the border. The second hole has the same radius and is located on the discontinuity parabola of the first hole, at a distance $d=0.2$. The reference magnetization is calculated on the same sample with only the first and the second holes. The third hole has a radius $R=0.05$ and is located at a constant distance  $d=0.2$ from the second hole. The applied magnetic field is $H_a=0.6~J_c$. Again, the maximum magnetization is obtained when the center of the third hole lies on the discontinuity parabola of the second one.} \label{Threeholes}
\end{figure}

Figure~\ref{Threeholes} shows the magnetization drop, $\Delta M/|M_2|$, as a function of the angular position of the third hole; the inset shows the configuration of the holes. The center of the third hole can be located in three different regions: region $A$, a region that is  not affected by holes $1$ and $2$, and regions $B$ and $C$, that respectively correspond to the influence regions of hole $2$ and hole $1$. Consider first that hole $3$ lies in region $A$. As $\theta$ increases, the magnetization drop, $\Delta M/|M_2|$, decays because the distance of hole $3$ from the border decreases (such a behaviour was already observed with the two-holes pattern). The magnetization drop is minimum when the center of hole $3$ lies on the discontinuity parabola of the hole $2$, separating regions $A$ and $B$. If $\theta$ increases further, hole $3$ enters region B, and $\Delta M/|M_2|$ increases again, to reach a maximum when the center of hole $3$ reaches the remnant parabola of hole $1$ (this line does not appear as a discontinuity line in the current line distribution). Then, $\Delta M/|M_2|$ decreases until hole $3$ reaches the boundary between regions $B$ and $C$ (this discontinuity line is a hyperbola). As it continues through region $C$, $\Delta M/|M_2|$ increases again. We can thus conclude that the magnetization drop is minimized each time the center of the hole is located on a discontinuity line. Note however that the values of $\Delta M/|M_2|$ on a minimum are not equal; the lowest value of $\Delta M/|M_2|$ is achieved on the boundary between regions $A$ and $B$.

\section{Influence of the type of lattices}
\label{s:lattice}

\subsection{Sample with a semi-infinite cross section}

\begin{figure}[b]
\center
\includegraphics[width=7cm]{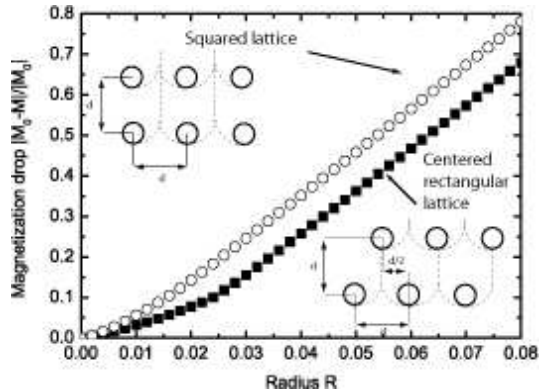}\caption{Magnetization drop (as compared to a sample with the same geometry and without holes), as a function of the hole radius in a sample with a semi-infinite cross-section (limited to a square of unit length) with two different lattice configurations. The open circle symbols correspond to a squared lattice with a lattice constant $d=0.2$ and the filled squared symbols refer to a centered rectangular lattice with the same lattice constant. The insets show the lattices. The number of holes for a given radius is the same in each lattice. The applied magnetic field is $H_a=0.6~J_c$. The centered rectangular lattice produces the largest magnetization.}\label{LatticeSI}
\end{figure}

Consider applying the results of the previous sections to construct a lattice containing many holes. Let us first compare the magnetization for two lattices. The first lattice is a squared lattice, where a line of equidistant holes (separated by a distance $d$) parallel to the external boundary is reproduced periodically every distance $d$ in the direction perpendicular to the border. The second lattice is obtained from the squared lattice by shifting every other row by half the length of a unit cell, leading to a particular realization of a centered rectangular lattice. The two lattices are represented in the insets of Figure~\ref{LatticeSI}. Since we are working with semi-infinite cross sections, we work in a square of  unit length, which we arbitrarily choose to contain five rows of holes. The holes have a common radius, $R=0.05$, and the lattice constant is fixed at $d=0.2$. Note that the hole density is equal for the two lattices.

The magnetization drop is defined as 
\begin{eqnarray}
\frac{\Delta M}{|M_0|}=\frac{|M_0-M|}{|M_0|},
\end{eqnarray}
where $M$ is the magnetization of the sample and $M_0$ stands for the magnetization for a sample with the same geometry but without holes. The applied field is carefully chosen to be $H_a = 0.6~J_c$ so that the flux front stays within the square of unit length. $\Delta M/|M_0|$ is plotted as a function of the hole radius in Figure~\ref{LatticeSI}. We find that the centered rectangular lattice produces a larger magnetization than the squared lattice. Although not shown, we checked that this result is independent of the hole separation distance $d$. This result naturally follows from the conclusions of the previous sections:  in the centered rectangular lattice, the holes are located on the discontinuity parabola of the neighbouring holes and the magnetization is maximized.

\subsection{Sample with a circular cross section}

Consider next infinitely long samples with a circular cross section. These samples have a geometry which is more realistic for bulk HTS applications. The Bean model in infinitely long geometries describes well the magnetic properties in the median plane of a cylinder with a finite height, provided its height $h$ is large with respect to its diameter $D$~\cite{Sam,Brandt}.

\begin{figure}[b]
\center
\includegraphics[width=7cm]{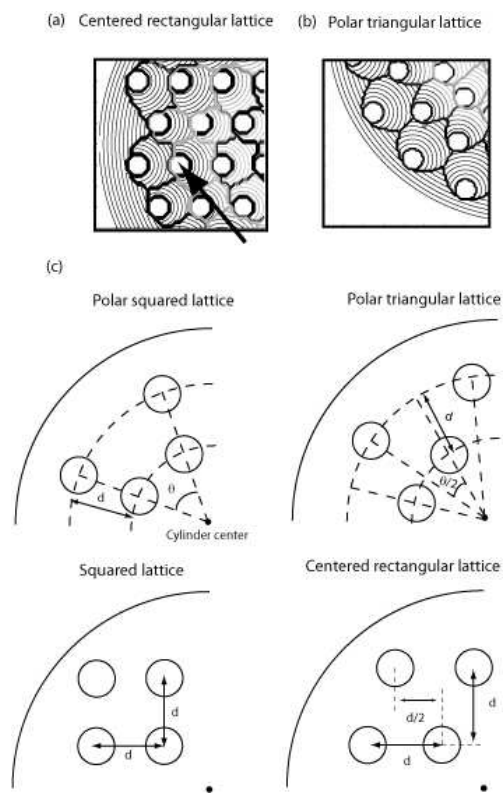}\caption{(a) Simulation of the current lines in a sample with a circular cross section and a centered rectangular hole lattice. The arrow indicates a hole which is not located on a discontinuity line. (b) Simulation of the current lines in a sample with a circular cross section and a polar triangular hole lattice. (c)  Lattice pattern under consideration in samples with a circular cross section.}\label{Polar}
\end{figure}

We found earlier that in a centered rectangular lattice, the holes were placed on the discontinuity lines of the neighbouring holes. This placement helped increasing the magnetization. However, this result is no longer correct for circular cross sections, because of the flux front geometry. The flux front is now circular, and as flux penetrates the system, the critical currents flow around concentric circular trajectories. Such a geometry is not compatible with the symmetry imposed by a centered rectangular lattice. The current lines for a centered rectangular hole pattern are represented in Figure~\ref{Polar}-(a). One can observe for instance that the hole indicated by the arrow is not located on a discontinuity line.

We can construct another lattice, that uses the circular shape of current lines and places the holes on discontinuity lines. Figure \ref{Polar}-(b) shows such a realization, which we name a ``polar triangular lattice''. The holes are positioned on concentric layers separated by a distance $d$. Inside each layer, the hole have a common angular separation. Furthermore, the holes are shifted every other layer by half their angular separation. This ensures that the holes are located on discontinuity lines. 

\begin{figure}[t]
\center
\includegraphics[width=7cm]{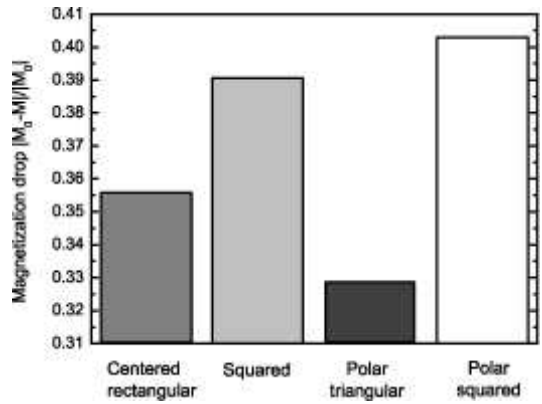}\caption{Magnetization drop in a sample with a circular cross section (unit radius) and sixty holes arranged in the four lattices presented in Figure~\ref{Polar}-(c). The holes have a radius $R=0.05~a$, the lattice constant is fixed at $d=0.2~a$, where $a$ is the radius of the cylinder. The angular separation in the polar lattice types is $20^\circ$. The sample is fully penetrated, $H_a=J_c~a$. The sample with the polar triangular hole lattice shows the smallest magnetization drop.}\label{LatticeCyl}
\end{figure}

By comparison with the squared lattice, one could also define a polar squared lattice where the hole angular position is not shifted from one layer to the next. The polar squared lattice, the polar triangular lattice, the squared lattice and the centered rectangular lattice are represented in a sample with a circular cross section in Figure~\ref{Polar}-(c). Each lattice contains sixty holes with  a radius $R = 0.05~a$, where $a$ is the radius of the cylinder. The lattice constant is fixed to $d=0.2~a$ and, for the polar lattices, the angular separation within a layer is fixed to $\theta=20^\circ$.  The corresponding magnetization drops are shown in Figure~\ref{LatticeCyl}. The applied field is such that the cylinder is fully penetrated, $H_a=J_c~a$. The reference magnetization $M_0$ is calculated for a sample with the same geometry and without holes. We thus find that the sample with the polar triangular hole lattice, which aligns holes of each layer on the discontinuity lines produced by the previous layers, has the smallest magnetization drop. According to the arguments of section~\ref{s:model}, this lattice will also have the highest trapped field.

The results of this study are based on the neglect of demagnetization effects and on the assumption that the critical current density is independent of the magnetic field strength. However, it is worth mentioning that, under the hypothesis of a constant critical current density, the remnant magnetization per unit volume is not influenced by demagnetization effects. Therefore, the result produced by the Bean model is also valid for a cylinder with a finite height, as already observed for bulk cylinders in Reference~\cite{Navau}. Thus, the conclusions drawn about the maximum magnetic flux that can be trapped remain applicable for cylinders of finite height.

\section{Conclusions}
\label{s:conclusions}

The magnetization drop induced by the removal of superconducting material in drilled samples has been studied numerically for different hole arrangements. We have developed an algorithm which calculates the magnetization in the critical state of infinitely long samples with an arbitrary hole pattern. The main principle of this algorithm lies on the shortest travel path for the flux front to reach a given point in the cross section. The algorithm successfully reproduces the discontinuity parabola attached to a single hole. The dependence of the hole radius on the magnetization drop of a sample with one hole indicates that the loss in magnetization scales neither with the surface of the hole ($\propto R^2$), nor with the surface of its parabolic region of influence ($\propto \sqrt{R}$), but as a surface of intermediate size, that is as $R^{3/2}$. From the simulations of samples with two and three holes, we have shown that in order to maximize the magnetization, the holes should always be located on discontinuity lines of their neighbours. The optimal lattice arrangement aligns the holes on the discontinuity lines and depends on the sample cross-section: we obtained the largest magnetization with a triangular hole lattice for samples with a semi-infinite cross section, and with a polar triangular lattice for samples with a circular cross section. 

\section{Acknowledgments}

G.L is grateful to the \textit{Fonds de la Recherche Scientifique (FRS-FNRS)} from Belgium for financial
support.

\appendix
\section{Analytical calculation of the magnetization drop in a sample with a semi-infinite cross section and one hole}
\label{appendix}

\begin{figure}[t]
 \centering
\includegraphics[width=7cm]{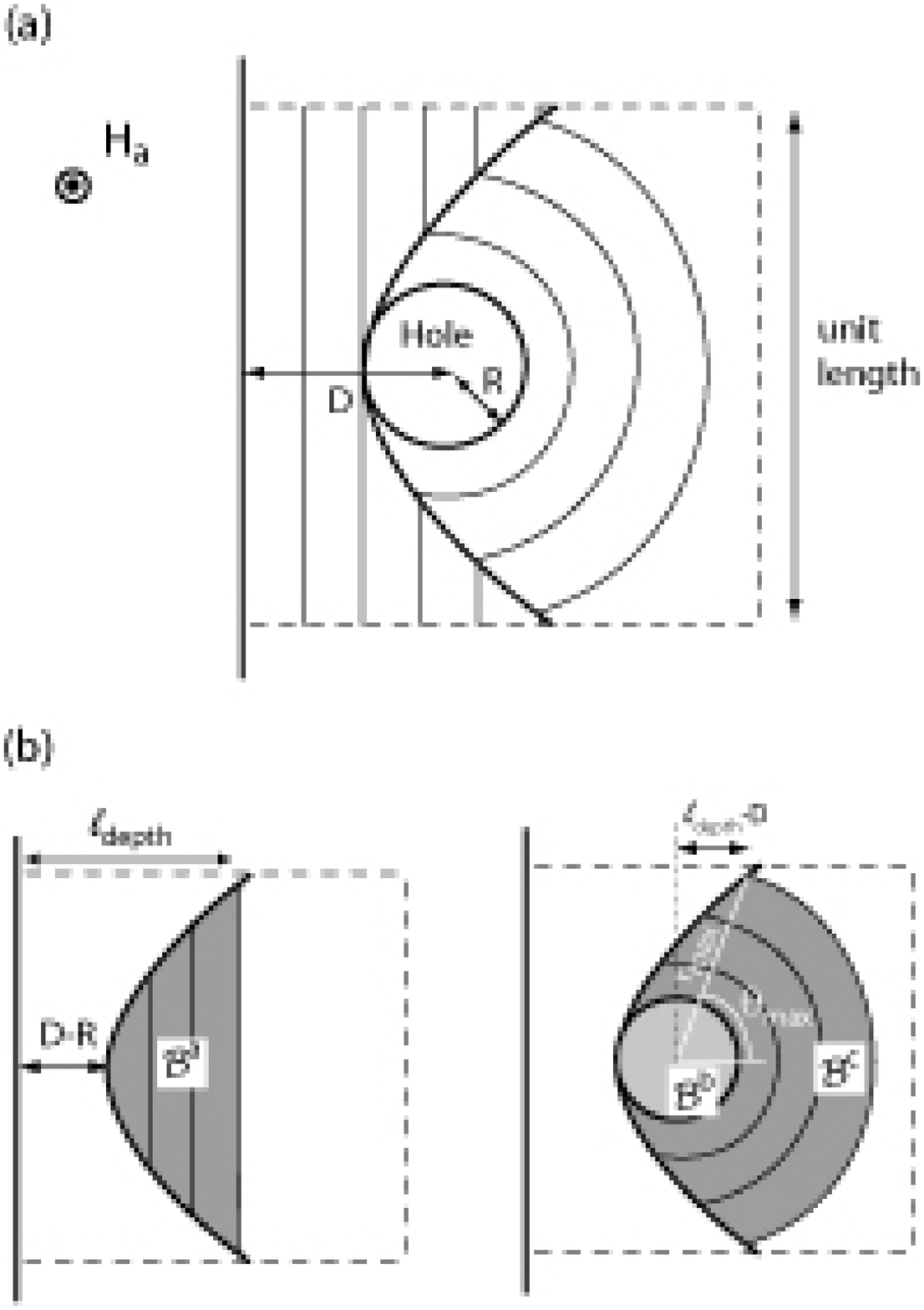}
\caption{\label{MagnDrop}{(a) Sketch of the current line in a sample with a semi-infinite cross section drilled by a hole of radius $R$ and located at a distance $D$ from the border. (b) Geometrical representation of the surface where we calculate magnetic flux for the magnetization difference $\Delta M$.}}
\end{figure}

Let us consider an infinitely long sample with a semi-infinite cross section and one hole of radius $R$, located at a distance $D$ from the border. The current line distribution is given by Figure~\ref{MagnDrop}-(a). The magnetization drop is defined as in the main text as $\Delta M/|M_0|=|M_0-M_1|/|M_0|$, where $M_0$ and $M_1$ respectively are the magnetization of the sample of unit surface, with and without a hole. The applied magnetic field oriented along the $z$-axis is given by $H_a$ and the penetration depth is $\ell= \ell_{\mathrm{depth}}$. The magnetic field has only a single component oriented along the $z$-axis. The magnetization is calculated as 
\begin{equation}
\label{magnetisation-calculation}
M=\frac{1}{\mu_0 S}\int_S B dS-H_a=\frac{1}{\mu_0}\mathcal{B}-H_a
\end{equation}
with $S=1$ is the cross section of a square of unit length. 

As the presence of the hole only modifies the flux front and the current lines inside the parabolic discontinuity line, the magnetization difference $\Delta M$ can be decomposed as 
\begin{equation} 
M_0-M_1=\frac{1}{\mu_0}\left(\mathcal{B}^a-(\mathcal{B}^b+\mathcal{B}^c)\right)
\end{equation} where $\mathcal{B}^{a,b,c}$ are respectively the average magnetic flux evaluated in the grey areas represented in Figure~\ref{MagnDrop}-(b). The three contributions are
\begin{equation}
\mathcal{B}^a=\frac{1}{S}\int_{D-R}^{\ell_\mathrm{depth}}\int_{-2\sqrt{Rx-Dx+R^2}}^{2\sqrt{Rx-Dx+R^2}}B(x)dy dx,
\end{equation}
where $B(x)=\mu_0(\ell_{\mathrm{depth}}-x)$, 
\begin{eqnarray}
\mathcal{B}^b & = & \frac{1}{S}\int_0^{2\pi}\int_0^R\mu_0(\ell_{\mathrm{depth}}-(D-R))rdrd\theta,\\ & = & \frac{\pi R^2}{S} \,(\ell_{\mathrm{depth}}-(D-R)),
\end{eqnarray}
and
\begin{eqnarray}
\mathcal{B}^c & = & \frac{2}{S}\left\{\int_{0}^{\theta_{max}}\int_{R}^{r_{max}}B(r)rdrd\theta+\int_{\theta_{max}}^{\pi}\int_{R}^{r_{parabola}}B(r)rdrd\theta\right\}, \label{Bc}
\end{eqnarray}
where $B(r)=\mu_0(\ell_{\mathrm{depth}}-r - D + 2 R)$. The integrals in the right side are carried in polar coordinates, with the origin fixed at the center of the hole. In these coordinates, the parabola equation is given as
\begin{equation}
r_{parabola}=\frac{2R}{1-\cos\theta}.
\end{equation}
The flux front intersects the parabola at an angle $\theta_\mathrm{max}$ that is defined by 
\begin{equation}
r_{max}=\frac{2R}{1-\cos\theta_{max}}
\end{equation}
with
\begin{equation}
r_{max}=\ell_\mathrm{depth}-D+2R.
\end{equation}

Carying the angular integrals (\ref{Bc}), we arrive at
\begin{eqnarray}
\mathcal{B}^c & = & \frac{2\mu_0}{S}\left\{\int_{R}^{r_{max}}\theta_{max}(\ell_{\mathrm{depth}}-r-D+2R)rdr \right. \\
 & & \left.   +\mu_0\int_{R}^{r_{parabola}}\left(\pi-\theta_{max}\right)(\ell_{\mathrm{depth}}-r - D + 2 R)rdr \right\} 
\end{eqnarray}
The integral over $r$ is computed numerically. 

The series expansion around $R=0$ of $\Delta M=|M_0-M_1|$ yields
\begin{equation}
\Delta M=\frac{J_c}{S} \left(\frac{32}{9}\sqrt{R^3(\ell_\mathrm{depth}-D)^3}+O(R^{5/2})\right)
\end{equation}

\end{document}